\documentclass[
12pt,
reprint,
showpacs,preprintnumbers,
showkeys,
amsmath,amssymb,
aps,
prb,
]{revtex4-1}
\usepackage{graphicx}
\usepackage{dcolumn}
\usepackage{bm}
\usepackage{color}
\usepackage{hyperref}

\newcommand{\LIA}{\lambda_{\rm I}^{\rm A}}
\newcommand{\LIB}{\lambda_{\rm I}^{\rm B}}
\newcommand{\LR}{\lambda_{\rm R}}

\begin{document}

\title{Theory of electronic and spin-orbit proximity effects in graphene on Cu(111)}

\author{Tobias \surname{Frank}}
 \email[Emails to: ]{tobias1.frank@physik.uni-regensburg.de}
\author{Martin \surname{Gmitra}}
\author{Jaroslav \surname{Fabian}}

\affiliation{%
 Institute for Theoretical Physics, University of Regensburg,\\
 93040 Regensburg, Germany
 }%

\date{\today}

\begin{abstract}
We study orbital and spin-orbit proximity effects in graphene adsorbed to the Cu(111) surface by 
means of density functional theory (DFT). The proximity effects are caused mainly by the 
hybridization of graphene $\pi$ and copper $d$ orbitals. Our electronic structure calculations agree 
well with the experimentally observed features.
We carry out a graphene\textendash Cu(111) distance dependent study to obtain proximity orbital and 
spin-orbit coupling parameters, by fitting the DFT results to a robust low energy model Hamiltonian. 
We find a strong distance dependence of the Rashba and intrinsic proximity induced 
spin-orbit coupling parameters, which are in the meV and hundreds of $\mu$eV range, respectively,
for experimentally relevant distances. The Dirac spectrum of graphene also exhibits 
a proximity orbital gap, of about 20 meV. Furthermore, we find a band inversion within the graphene 
states accompanied by a reordering of spin and pseudospin states, when graphene is pressed towards 
copper.
\end{abstract}

\pacs{71.15.Mb, 73.22.Pr}
\keywords{DFT; graphene; Cu(111); surface; SOC}
\maketitle


\section{\label{sec:intr} Introduction}

Copper is an important material for graphene. 
Graphene\textendash copper junctions are often encountered in technological applications.\cite{Giovannetti2008,Khomyakov2009}
For example, graphene can be used to seal a copper surface to preserve its excellent plasmonic characteristics.\cite{Kravets2014} The 
growth of graphene via CVD by the deposition of CH$_4$ on copper surfaces is 
amongst the most popular techniques to obtain large (poly)crystalline graphene. 
\cite{Gao2010} Even single layer graphene grains of millimeter size as well as 
pyramid-like bi- and trilayer graphene, hexagonal onion ring-like graphene 
grains can be grown on copper.\cite{Yan2014,Boyd2015} Important to our study,
graphene produced on a copper surface exhibits a giant spin Hall effect, lilkely
due to residual cupper adatoms and ad-clusters.\cite{Balakrishnan2014}

Experimentally, graphene on the Cu(111) surface has been well studied 
by means of angle-resolved photoemission spectroscopy (ARPES) \cite{Avila2013,Jeon2013,Marsden2013,Shikin2013,Varykhalov2010,Vita2014, 
Walter2011} and scanning tunneling microscopy (STM).\cite{Gao2010} The linear dispersion of graphene
is found to be preserved. ARPES measurements of the graphene\textendash Cu(111) system find, that graphene 
is getting electron doped,\cite{Walter2011} leading to a shift of the Dirac energy $E_\text{D}$, 
which we define as the average energy of the graphene $\pi$ state energies at 
K with respect to the Fermi energy $E_\text{F}$. Typically, $E_\text{D}$ is about $-0.3$~eV with respect to the Fermi energy.\cite{Walter2011} The top of the $d$ band edge of copper begins at 
$-2$~eV below the Fermi level. It is observed that a gap opens within the Dirac 
cone of graphene of about 50--180~meV.\cite{Avila2013,Marsden2013,Shikin2013,Varykhalov2010,Vita2014,Walter2011}

Spin-orbit coupling (SOC) effects in graphene on selected metal substrates were studied 
theoretically \cite{Abdelouahed2010, Li2011} and experimentally \cite{Rader2009, Marchenko2012} 
and it was 
noticed that substrates can induce sizeable spin-orbit effects important for spintronics applications.~\cite{Zutic2004,Han2014}
Spin resolved ARPES experiments\cite{Shikin2013} focused on the spin-orbit 
effects introduced by metallic surfaces in graphene, investigating the role of 
the atomic number of the substrate. It was found that the states of graphene can 
be split due to Rashba spin-orbit coupling by up to 100~meV in the case of Au 
and 10~meV in the case of Ni,\cite{Li2011} respectively. Copper substrate induced 
spin-orbit splittings in graphene are expected to be substantially smaller.\cite{Shikin2013} They were measured at a temperature of 40 
K, which gives a resolution limit and also the upper bound for the spin-orbit effects of 3.4 meV. The mechanism introducing the 
spin-orbit interaction was identified to be the hybridization between substrate $d$
and graphene $\pi$ states. \cite{Shikin2013} Our present work agrees with this conclusion, and
predicts the values of the Rashba splitting to be about 2 meV for a reasonable distance 
between graphene and copper, just below the stated experimental resolution of Ref. \onlinecite{Shikin2013}.

Crucial to obtain accurate graphene\textendash metal distances is to consider van der Waals interactions.
It was found\cite{Olsen2013, Vanin2010,Andersen2012}, that the dispersive long-range interactions 
play an important role in binding, yielding graphene\textendash copper distances of 2.91 to 
3.58 \AA.

Here, we focus on hybridization and proximity effects by means of DFT calculations.
By the application of an effective Hubbard $U$,\cite{Anisimov1991}
which corrects for self interaction errors, we achieve a good agreement with experiment in terms of the emission spectra and the band structure features.
We carry out an analysis of the orbital composition of the band structure, giving us hints for a model
Hamiltonian including spin-orbit interactions, which can be used to describe graphene
in combination with many other materials that yield a $C_{3v}$ or higher symmetric system. We then fit
the DFT data to the model Hamiltonian and extract parameters such as the induced gap
as well as spin-orbit coupling values. As the graphene\textendash copper distance is not exactly known
experimentally,  and there is still a theoretical uncertainty in determining its magnitude, we carry out a distance-dependent study.

Our main finding is a strong graphene\textendash Cu(111) distance-dependent spin-orbit coupling introduced in the graphene states.
We use a model Hamiltonian to describe those states, for which we observe a Rashba spin-orbit coupling parameter which
reaches values of meVs, while being absent in pristine graphene. The proximity induced intrinsic SOC is in the hundreds of $\mu$eV range,
a factor of ten larger than in pristine graphene.
We also observe a closing of the induced gap for a graphene\textendash copper distance of 2.4~\AA. This is accompanied
by a peculiar reordering of spin and pseudospin states associated with a gap inversion at small distances.

The paper is organized as follows. Sec. \ref{sec:comp_meth} deals with the computational methods used.
Geometrical structure modeling is described in Sec. \ref{sec:unit_cell}. In Sec. \ref{sec:methods_choice} we carry out the analysis
of the band structure. In Sec. \ref{sec:mod_ham} we introduce our model Hamiltonian and fit it to the ab-initio data. Finally, in Sec. \ref{sec:dist_stud} we present our graphene\textendash copper distance dependent study
with a discussion of the proximity induced effects.

\section{\label{sec:comp_meth} Computation methods}
We used DFT implemented in the plane-wave code \textsc{quantum espresso}.\cite{Giannozzi2009}
The calculations were performed at a $k$ point sampling of $40 
\times 40$ if not indicated otherwise. A slab geometry was applied, 
where we added a minimum of 15~\AA\, of vacuum around the structure in $z$ 
direction. We used the Kresse-Joubert ultrasoft (relativistic) PBE\cite{Perdew1996} projector augmented wave 
pseudopotentials.\cite{Kresse1999} The plane wave energy cutoff was set to 40~Ry and the charge 
density cutoff to 320~Ry to ensure converged results. Van der Waals interactions were taken into 
account using the empirical method of Grimme.\cite{Grimme2006} To cross check 
spin-orbit coupling calculations we also employed the all electron, full potential linearized 
augmented plane wave code \textsc{wien2k}.\cite{Blaha2014} We found that spin-orbit coupling
splittings were differing at most by 10\%. For processing our 
distance studies the atomic simulation environment (ASE)\cite{Bahn2002} was 
used. Hellmann-Feynman forces in relaxed structures were decreased until they 
were smaller than $0.001~\text{Ry}/a_0$. Calculations for graphene on Cu(111) 
included calculations adding the Hubbard $U$ correction.\cite{Cococcioni2005} 

\section{\label{sec:graphene-cu111}Graphene C\lowercase{u}(111) study}
\subsection{\label{sec:unit_cell} Choice of unit cell}
\begin{figure}
 \includegraphics[width=0.4\textwidth]{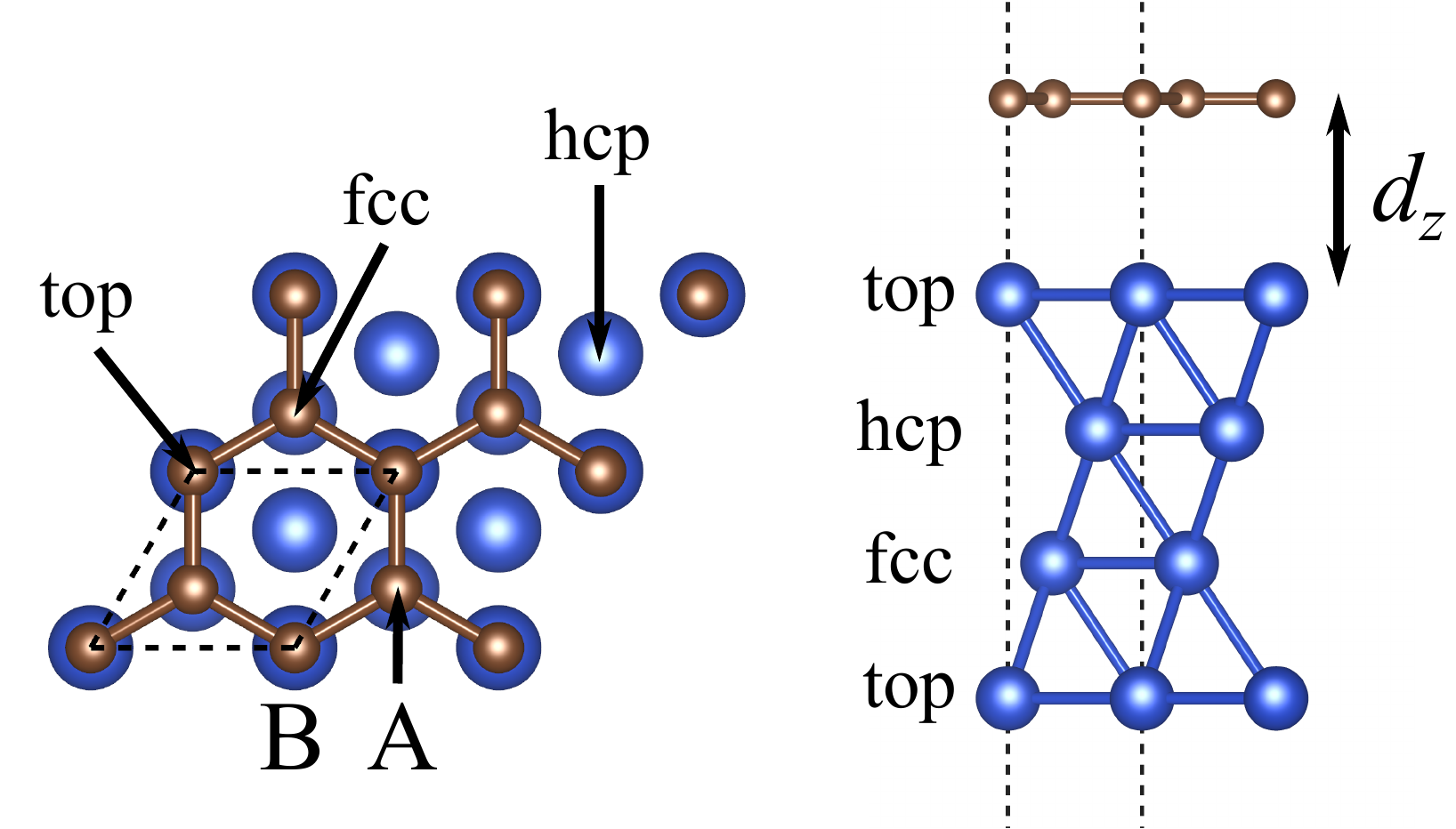}
 \caption{(color online) \label{fig:unitcell}Structure: Top and side views of 
the unit cell, which is indicated as black dashed lines, repeated twice in each 
lateral direction. Blue (large) spheres indicate the copper atoms, brown (small) 
spheres the carbon atoms. The sublattice is depicted by labels A and B. The 
copper layers are labeled by top, hcp and fcc, which also tag the adsorption 
positions.}
\end{figure}

The mismatch between Cu(111)'s surface lattice constant of $3.61/\sqrt{2}$~\AA\,\cite{Straumanis1969} 
and graphene's lattice constant of 2.46~\AA\, is 3.8\%. STM experiments\cite{Gao2010} observe regions with different 
moir\'{e} structures; the most observed one (30\%) is a commensurate lattice configuration with 
a periodicity of 66~\AA. Another experiment\cite{Avila2013} found that 60\% of graphene grains on Cu(111) 
are preferentially rotated by $3^\circ$ with respect to the substrate. To account for the lattice mismatch, 
one would have to chose a unit cell, which is computationally very demanding, containing hundreds of atoms. 
We set the lattice constant of copper to be compatible with the experimental graphene lattice constant, 
to describe graphene as realistically as possible, following Ref.~\onlinecite{Khomyakov2009}. 
A supporting fact to use graphene's lattice constant is that graphene does not chemically bind to copper and its strong in-plane 
$\sigma$ bonds remain intact.

In the transverse direction to the Cu(111) surface one distinguishes three non-equivalent Cu planes.
We label the planes from the surface towards bulk as top, hcp and fcc, see Fig.~\ref{fig:unitcell}.
We tested three different commensurate configurations named according to which Cu layer the carbon 
atoms sit over. This gives rise to three possible graphene physisorbed positions named as top-fcc,
top-hcp, and fcc-hcp configuration.\cite{Khomyakov2009}
In Fig.~\ref{fig:unitcell} we show the top-fcc configuration, where one carbon atom (say from sublattice B) 
is on top of a copper atom of the top layer, while the other carbon (from sublattice A) is over the 
fcc Cu layer.

In general, the graphene sublattices have different environments. This breaks the sublattice symmetry of 
graphene and results in sublattice resolved spin-orbit coupling effects.\cite{Gmitra2013, Gmitra2015}
To simulate a copper surface we used four layers of copper. We checked that the physics of the graphene low energy states 
does not change upon increasing the number of layers. In addition, we found good agreement of the band structure with
experiment.\cite{Avila2013,Jeon2013,Marsden2013,Shikin2013,Varykhalov2010,Vita2014,Walter2011}

In our studies we first relaxed the copper slab \emph{alone} without van der Waals corrections and then fixed 
its degrees of freedom and let just the carbon atoms relax in $z$ direction including empirical 
van der Waals corrections.\cite{Grimme2006} To start with, the copper slab is strained in the $xy$ plane such that its surface lattice constant $a_\text{Cu}/\sqrt{2}$ 
is the same as the experimental graphene lattice constant of 2.46~\AA\, yielding an effective bulk lattice 
constant of $a_\text{Cu} = 3.48$~\AA. This represents a compression of the copper slab by 3.8\% with respect
to the bulk value of 3.61~\AA.\cite{Straumanis1969} After letting the copper slab relax in
$z$ direction, the distance of copper atoms from plane to plane was 2.59~\AA, corresponding 
to an expansion of 1.7\% compared to bulk copper. This compensates to some extent for the compression in the $xy$ plane.

Comparing the top-fcc with the other commensurate configurations top-hcp and fcc-hcp we found 
slightly different graphene\textendash Cu(111) distances $d_z$ of 3.10~\AA, 3.11~\AA, and 3.12~\AA, respectively. The 
corrugation of the carbon atoms in $z$ direction is less than $10^{-3}$~\AA, expressing the 
weak nature of binding. The lowest energetic configuration is the top-fcc arrangement, followed 
by the top-hcp, which is only 2.3~meV higher in energy per unit cell. The highest one in total energy with 12.3~meV 
compared to top-fcc is fcc-hcp, where the nearest copper atom sits within the carbon ring. Therefore 
in the following study we consider the top-fcc configuration.

\subsection{\label{sec:methods_choice}Choice of methods and electronic structure}

\begin{figure}
 \includegraphics[width=0.48\textwidth]{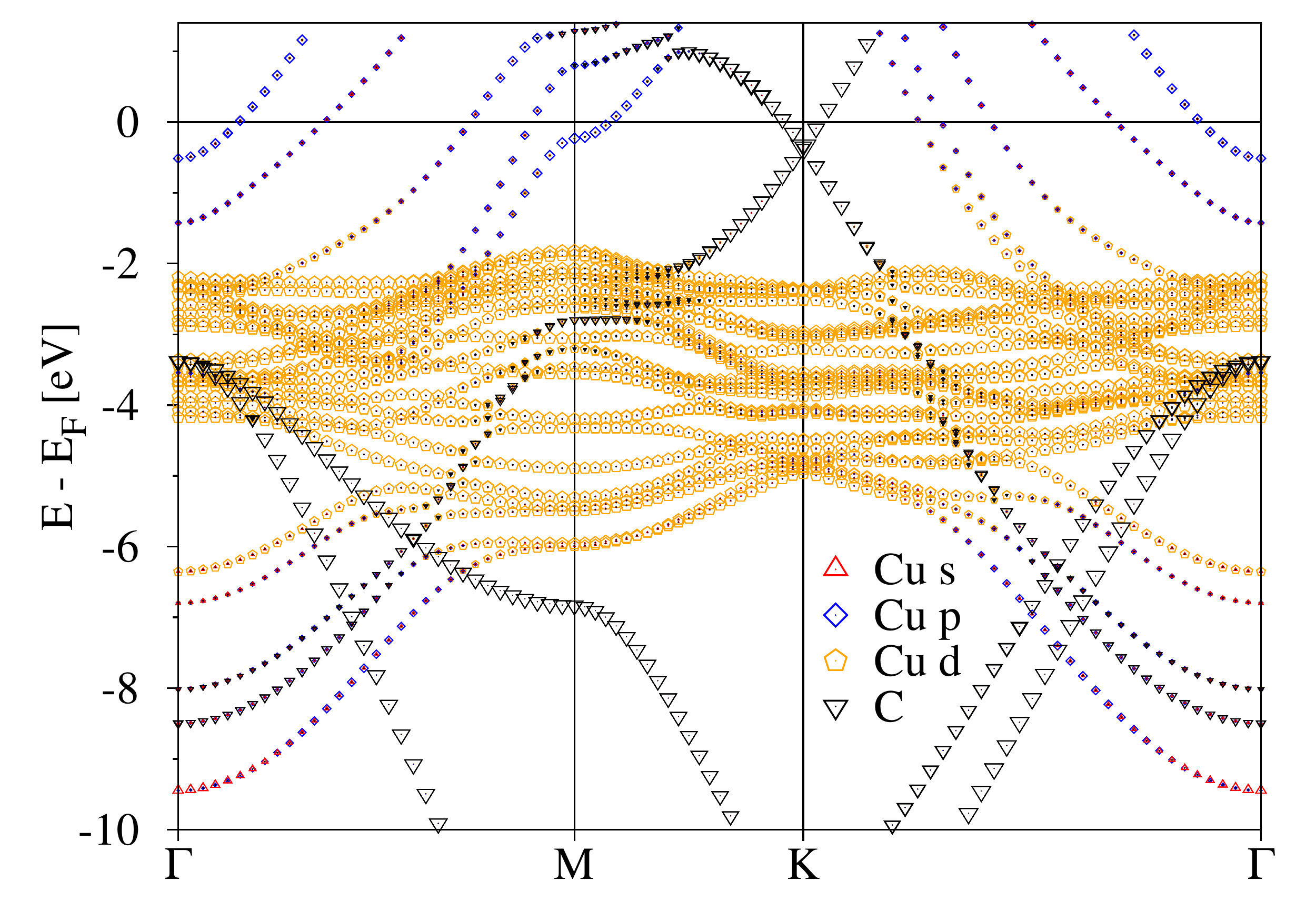}
 \caption{(color online) \label{fig:fatband}
Calculated electronic structure of graphene/Cu(111) slab. The graphene distance from Cu(111) surface is $3.09$~\AA. 
The overlaying symbols indicate orbital resolved contributions to the eigenvalues. 
Orange pentagons show Cu $d$ bands, red upward pointing triangles represent Cu $s$ states, 
blue squares show Cu $p$ states and black downward pointing triangles indicate graphene states.}
\end{figure}

The orbital resolved electronic structure of graphene on Cu(111) is shown in Fig. \ref{fig:fatband}
for DFT+$U$ with an effective $U=1$~eV \cite{Cococcioni2005} acting on the Cu 3$d$ electrons
for a copper\textendash graphene distance of 3.09~\AA.
It can be seen, that the Dirac cone structure is preserved for energies higher 
than $-2$~eV. Below this energy region the graphene $\pi$ states hybridize with the copper $d$ states. 
This can be seen by the avoided crossings if one follows the $\pi$ band towards the $\Gamma$ point at $-8.5$~eV. On this way,
at $-6$~eV the $\pi$ states branch and strongly hybridize with a copper band consisting of $p$ and $s$ states. Those are states
which are situated on the surfaces of the slab and whose degeneracy is broken due to the graphene potential.
The graphene $\sigma$ states starting from $-3.5$~eV at the $\Gamma$ point are mainly unaffected.
The copper $s$ and $p$ states are present in the energy region between $-9.5$~eV and $-6$~eV as well as from
$-2$~eV and upwards. The copper band structure obtained here is qualitatively in agreement with bulk fcc calculations. The position of the Fermi energy 
was converged for a dense sampling of the Brillouin zone.

There is charge transfer from the Cu(111) surface to graphene. As a result,
graphene gets $n$-doped.\cite{Khomyakov2009} The Dirac cone is shifted below the Fermi level by about 
$E_{\rm D}=0.3$~eV according to experiments.
We compared the effect of the relaxation of copper slabs in $z$-direction for the relaxed and non-relaxed (bulk lattice constant of $3.48$~\AA)
cases on the doping of graphene. For non-relaxed (compressed) copper slabs the electron doping of graphene was 
significantly higher than for relaxed slabs due to the higher kinetic energy in non-relaxed copper. For relaxed copper 
slabs the Dirac energy shift is comparable to experiment,\cite{Gao2010} being 350~meV.

To account for correlation effects, we applied a Hubbard $U$ correction of 1~eV to the copper $d$ states. 
In this way we match the onset of the completely filled copper $d$ levels, which show up at $-2$~eV below the Fermi energy in the ARPES 
experiments.\cite{Avila2013,Marsden2013,Shikin2013} The effect of the Hubbard $U$ correction 
is a rigid shift of the filled copper $d$ levels to lower energies without changing their band widths.
However, we see a strong dependence of the copper $d$ level energies on the compression of the copper slab, 
they are 1~eV higher in energy for the compressed than for the relaxed one. The proper position of the $d$ levels
is significant for the spin-orbit coupling induced proximity effect in the Dirac cone, as there can be larger hybridization,
when the $d$ levels are closer to the states of interest.

All in all we find a good agreement of the band structure with experiment.\cite{Avila2013} 
The only shortcoming is the description of the graphene gap, which is opening 
at the Dirac energy $E_\text{D}$. We find it to be 20~meV, which is lower than the 50 to 180~meV 
stated in experiments.\cite{Avila2013,Marsden2013,Shikin2013,Varykhalov2010,Vita2014,Walter2011} This deviation
could be due to the limitations of semilocal and local exchange-correlation functionals.

\subsection{\label{sec:mod_ham} Model Hamiltonian}
As we demonstrated above, DFT+$U$ reasonably captures the electronic structure of graphene 
on the Cu(111) surface. Now we use the first-principles calculations to predict proximity
induced effects of the copper surface on the spin-orbit coupling in graphene. For this purpose
we study a Hamiltonian describing the low energy $\pi$ states of graphene on Cu(111).
The Hamiltonian ${\cal H}={\cal H}_{\rm orb}+{\cal H}_{\rm so}$ contains orbital and spin-orbit coupling parts
and describes graphene whose symmetry point group is lowered from $D_\text{6h}$ (pure graphene) to $C_\text{3v}$. Such a Hamiltonian was introduced already in the context
of hydrogenated graphene \cite{Gmitra2013} in which the pseudospin symmetry
gets broken explicitly by hydrogenation, but it was also found useful in graphene whose pseudospin is broken implicitly only, by placing graphene on incomensurate lattices such as MoS$_2$.~\cite{Gmitra2015} In our case the pseudospin symmetry is broken 
explicitly as the pseudospin state is well defined but the two sublattices experience a different
orbital environment, see Fig. \ref{fig:unitcell}. This proximity Hamiltonian has the form,
\begin{equation}
 {\cal H}_\text{orb} = \hbar v_{\rm F}(\kappa \sigma_x k_x + \sigma_y k_y) + \Delta \sigma_z s_0\,,
 \label{eq:Horbital}
\end{equation}
and
\begin{align}
{\cal H}_\text{so} = &\lambda_\text{I}^\text{A}\left[\left( \sigma_z + \sigma_0\right)/2\right]\kappa s_z + \lambda_\text{I}^\text{B}\left[\left( \sigma_z - \sigma_0\right)/2\right]\kappa s_z \label{eq:Hlambda}\\
 &+ \lambda_\text{R} \left( \kappa \sigma_x s_y - \sigma_y s_x \right)\,, \label{eq:Hrashba}
\end{align}
where $v_{\rm F}$ is the Fermi velocity and $\kappa=1(-1)$ labels the valley degree of freedom.
$k_x$ and $k_y$ are the Cartesian components of the electron wave vector
measured from K(K$^\prime$), $\sigma_x$ and $\sigma_y$ are the pseudospin Pauli
matrices acting on the two-dimensional vector space formed by the two 
triangular sublattices of graphene. The first term in ${\cal H}_{\rm orb}$
describes gapless Dirac states. The second term describes the effective orbital
hybridization energy, which acts as a staggered potential on sublattices A and B, where $\sigma_z$ is the pseudospin Pauli matrix and $s_0$ is the
unit matrix in spin space. This Hamiltonian term leads to an orbital proximity induced gap in the Dirac spectrum of $2\Delta$. This gap is still present
even when spin-orbit coupling is turned off. A consequence
of the pseudospin inversion asymmetry is the sublattice-resolved intrinsic
spin-orbit coupling. As intrinsic spin-orbit coupling is a next-nearest
neighbor hopping, it acts solely on a given sublattice. We describe it
with parameters $\LIA$ and $\LIB$ for sublattice A and B, respectively.
We denote by $s_z$ the spin Pauli matrix and by $\sigma_0$ the unit matrix
acting on the pseudospin space. If $\LIA \not= \LIB$, the spin degeneracy
gets lifted already by this intrinsic term, reflecting the loss of
space inversion symmetry. The space inversion asymmetry itself gives rise
to Rashba type spin-orbit coupling whose strength is measured by $\LR$, which
is a nearest-neighbor spin-flip hopping, contributing further to the
spin splitting of the low energy bands.

The four eigenvalues of the model Hamiltonian at the K point ($k=0$) read
\begin{align*}
\varepsilon_4 &= -\frac{1}{2}\lambda_{\rm I}^{+} + \sqrt{(\Delta - \frac{1}{2}\lambda_{\rm I}^{-})^2 + 4\lambda_\text{R}^2}, \\
\varepsilon_3 &= \Delta + \frac{1}{2}\left(\lambda_{\rm I}^{+}+\lambda_{\rm I}^{-}\right),  \\
\varepsilon_2 &= -\Delta + \frac{1}{2}\left(\lambda_{\rm I}^{+}-\lambda_{\rm I}^{-}\right),  \\
\varepsilon_1 &= -\frac{1}{2}\lambda_{\rm I}^{+} - \sqrt{(\Delta - \frac{1}{2}\lambda_{\rm I}^{-})^2 + 4\lambda_\text{R}^2},
\end{align*}
where $\lambda_\text{I}^{+} = \lambda_\text{I}^\text{A}+\lambda_\text{I}^\text{B}$ and 
$\lambda_\text{I}^{-} = \lambda_\text{I}^\text{A}-\lambda_\text{I}^\text{B}$ for compactness. 
We ordered the eigenvalues by decreasing energies, where we assumed 
$\Delta \gg \lambda_\text{R} \gg \lambda_\text{I}^\text{A},\, \lambda_\text{I}^\text{B}$. 
The eigenstates $\varepsilon_2$ and $\varepsilon_3$ always have spin-$z$ expectation values of $s_z=-1/2$ and 
$s_z=1/2$, and pseudospin-$z$ expectation values of $\sigma_z=-1/2$ and $\sigma_z=1/2$ and are localized on 
sublattice B and A, respectively. The eigenstates with $\varepsilon_1$ and $\varepsilon_4$ 
in general are mixtures of sublattices and spin directions, but have almost
$s_z \simeq 1/2$, $\sigma_z \simeq -1/2$ and $s_z \simeq -1/2$, $\sigma_z\simeq 1/2$ 
under the assumption that $\Delta \gg \lambda_\text{R} \gg \lambda_\text{I}^\text{A}$, 
$\lambda_\text{I}^\text{B}$. In the model Hamiltonian there are four unknown parameters.
To construct a set of independent equations we also take into account the spin-$z$ expectation value
for the first eigenstate denoted by $s_1^z$. The model parameters thus can be expressed as follows
\begin{align*}
\Delta &=  \frac{1}{4} \left( -\varepsilon_2 + \varepsilon_3 -2s^z_1 (\varepsilon_1 - \varepsilon_4 )\right),\\
\lambda_\text{I}^\text{A} &= \frac{1}{4} \left( -\varepsilon_1 + 2 \varepsilon_3 - \varepsilon_4 + 2 s^z_1(\varepsilon_1 - \varepsilon_4 )\right),\\
\lambda_\text{I}^\text{B} &= \frac{1}{4} \left( -\varepsilon_1 + 2 \varepsilon_2 - \varepsilon_4 - 2 s^z_1(\varepsilon_1 - \varepsilon_4 )\right),\\
\lambda_\text{R} &= \frac{1}{4} \left( \varepsilon_1 - \varepsilon_4\right) \sqrt{1 - 4(s^z_1)^2}.
\end{align*}
We note that special care has to be taken when associating the order of 
the DFT eigenvalues with respect to the model Hamiltonian eigenvalues. 
For every state we compared the sublattice localization and $s_z$ values for 
both the DFT and model calculations.

\begin{figure}
 \includegraphics[width=0.47\textwidth]{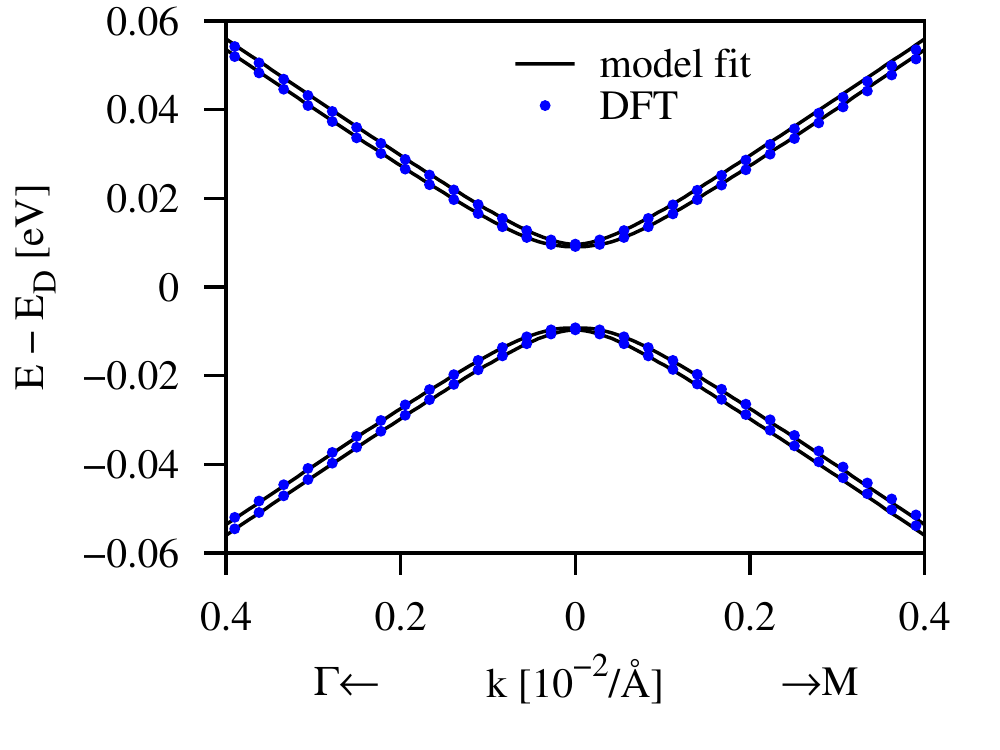}
 \caption{(color online) \label{fig:bands_fit} 
Calculated band structure around Dirac point. Comparison of DFT calculations 
with the model calculations for a graphene\textendash copper distance of 3.09~\AA. 
The energy is measured with respect to the Dirac energy $E_\text{D}$. The plot is centered at K ($k=0$) and its left part corresponds to the $k$
points pointing towards $\Gamma$ and the right part towards the M point. 
}
\end{figure}

In Fig.~\ref{fig:bands_fit} we compare low energy graphene bands calculated from DFT and model, for a distance of $d_z=3.09~{\textrm \AA}$. The fitted model parameters are 
$\Delta = 9.3\,\text{meV}$, $\lambda_\text{I}^\text{A} = -0.131$~meV,
$\lambda_\text{I}^\text{B} = 0.060$~meV, and $\lambda_\text{R} = 1.2\,\text{meV}$.
The proximity effects, both, the orbital and spin-orbit coupling ones are significant.
The hybridization gap $\Delta$ dominates the energy scale. It yields a gap value 
of $E_\text{gap} = \varepsilon_3 - \varepsilon_2 \approx 2 \Delta = 18.6$~meV. 
The Rashba spin-orbit coupling parameter of 1.2~meV indicates a very strong effect of the
space inversion asymmetry, which would correspond to a transverse electric field
of 240~V/nm for bare graphene.\cite{Gmitra2009} The intrinsic 
spin-orbit coupling parameters have opposite sign and their amplitudes are significantly enhanced in comparison to the tens of $\mu$eV 
in bare graphene.\cite{Gmitra2009}
The Fermi velocity is $v_\text{F} = 0.825 \cdot 
10^6~\text{m}/\text{s}$ (equivalent to a nearest neighbor hopping of 2.55~eV). 
We see that the band structure is isotropic in this range of $k$ points and the model description 
agrees very well with the DFT data. We observed a good agreement up to energies 
$\pm 0.1$~eV away from the Dirac energy. 

\begin{figure}
 \includegraphics[width=0.47\textwidth]{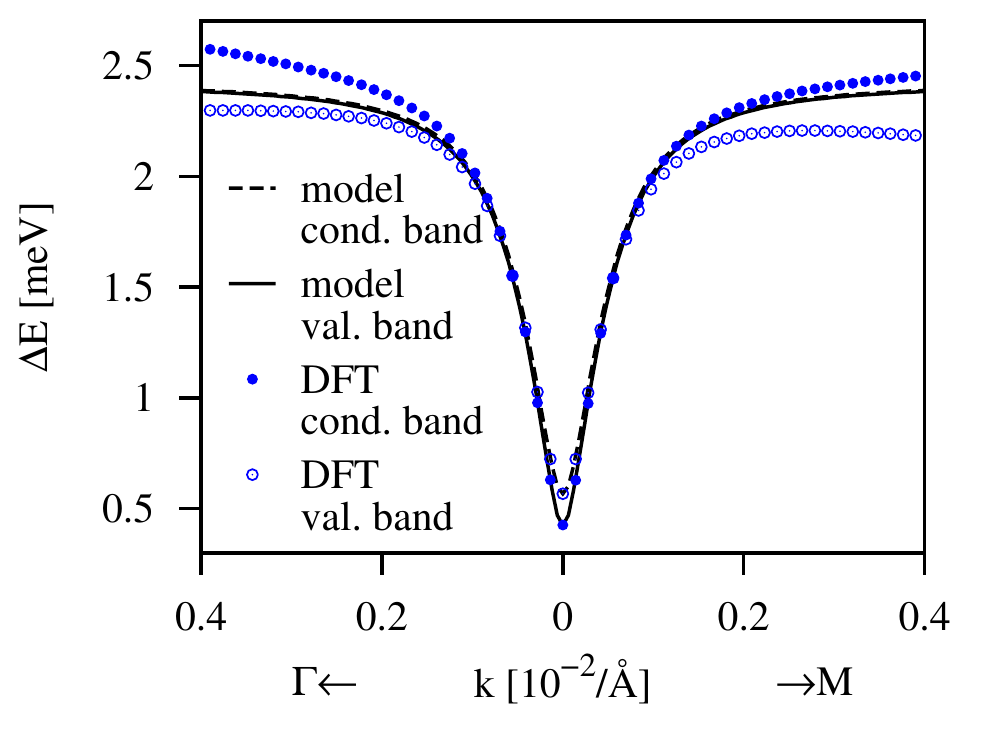}
 \caption{(color online) \label{fig:splittings_fit} 
Calculated spin splittings of the valence and conduction bands.
The DFT data is shown by symbols while the lines correspond to the model description.
The distance between graphene and copper is 3.09~\AA.
}
\end{figure}

We also compare the band spin splittings of the valence and conduction bands, 
see Fig.~\ref{fig:splittings_fit}. 
It can be seen, that by construction, the splittings at K are described exactly. 
The model reproduces very well the narrowing of the band splittings for 
$k$ points up to $0.1 \cdot 10^{-2}/{\textrm \AA}$ away from the K point even though only information
from the K point enters.
As the model does not include spin-orbit coupling terms dependent on $k$, 
both the valence and conduction band splittings from the model calculations
saturate at a common value for larger $k$ due to the Rashba SOC. 
To include $k$ dependent terms one needs to consider terms such as pseudospin inversion asymmetry 
(PIA)\cite{Gmitra2015,Gmitra2013,Irmer2015} which can capture the $k$ dependence of the
splittings.
In the DFT calculations we observed that the splittings for valence (conduction) bands 
increase (decrease) with larger distances from K as the interaction with 
copper $d$ levels increases (decreases) and the induced spin-orbit effects are 
stronger (weaker). 

\begin{figure*}
 \includegraphics[width=0.7\textwidth]{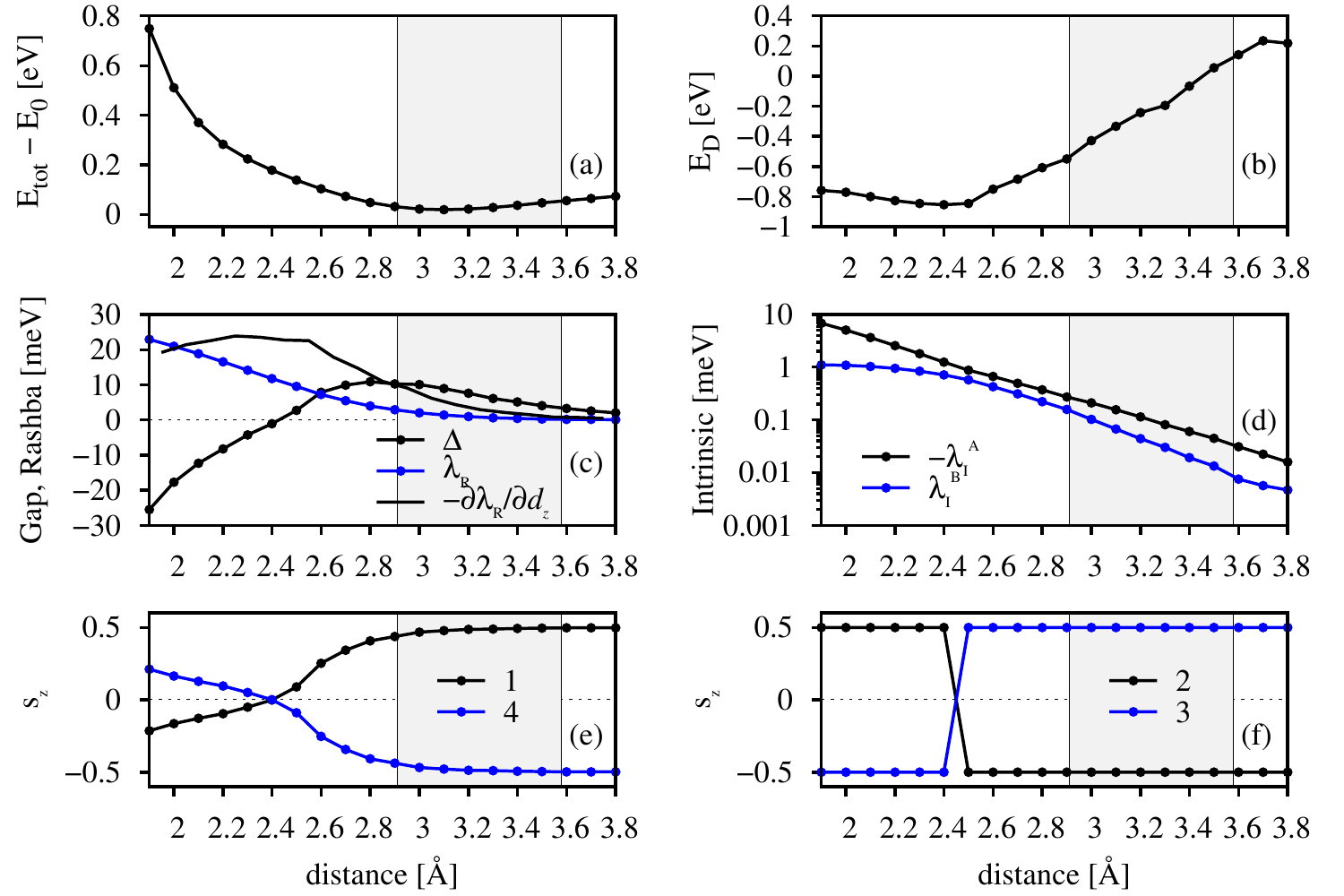}
 \caption{(color online) \label{fig:dist_study}
Calculations of low energy properties of graphene on Cu(111) surface as a function of distance, with a Hubbard $U$ of 1 eV used.
(a)~Total energy with respect to the minimal total energy at 3.09~\AA;
(b)~Dirac energy shift $E_\text{D}$ with respect to Fermi level;
(c)~proximity induced potential $\Delta$ and Rashba spin-orbit coupling parameter $\lambda_\text{R}$, as well as the derivative of $\lambda_\text{R}$;
(d)~intrinsic spin-orbit coupling parameters $\lambda_\text{I}^\text{A}$ and $\lambda_\text{I}^\text{B}$;
(e)~spin $s_z$ expectation values for the $\varepsilon_1$ and $\varepsilon_4$ graphene eigenvalues 
at the K point and (f)~for the $\varepsilon_2$ and $\varepsilon_3$ eigenvalues. The shaded region 
indicates predicted distances from other theoretical references.\cite{Olsen2013, 
Vanin2010,Andersen2012}}
\end{figure*}

\subsection{\label{sec:dist_stud} Distance study}

Standard DFT can not account for dispersive forces. Different methods dealing 
with van der Waals effects often yield inconsistent results\cite{Vanin2010,Olsen2013,Andersen2012} 
when trying to treat graphene on metal surfaces. Therefore, we conduct calculations of electronic 
properties for different graphene\textendash Cu(111) distances. We used the Hubbard 
correction\cite{Cococcioni2005} with 
$U=1$~eV for Cu $d$ electrons. The relative coordinates of the 
atoms within the copper slab and within graphene were fixed and the graphene\textendash copper distance $d_z$ was varied.
We apply the same analysis as in Sec.~\ref{sec:mod_ham} 
for each distance configuration $d_z$ and extract the total energy of the structure,
the Dirac energy shift $E_\textrm{D}$, the hybridization gap $\Delta$, the Rashba and intrinsic spin-orbit coupling
parameters as well as spin-$z$ expectation values of the graphene states at the K point.

In Fig.~\ref{fig:dist_study}(a) we show the total energy as a function of the graphene
distance $d_z$ from the Cu(111) surface. The curve is shifted with respect to the minimal total 
energy at the distance of 3.09~${\textrm \AA}$. The energy dependence has a rather shallow
minimum where the energy increases by just 0.5 eV when graphene is pushed to a distance of 2~\AA.

Fig.~\ref{fig:dist_study}(b) visualizes the shift of the Dirac energy $E_{\textrm D}$ with respect to the 
Fermi level. We see that graphene stays $n$-doped for distances smaller than 3.5 ${\textrm \AA}$, and the curve has two regimes. 
For larger distances down to 2.5~${\textrm\AA}$ there is a linear behavior with a positive slope,
the more graphene is pushed towards the Cu(111) surface, the more $n$-doped it gets.
For distances smaller as 2.5~${\textrm\AA}$ the slope reverses its sign and is more shallow. This means that 
there occurs a significant charge transfer from the copper slab to the graphene sheet, which 
saturates at smaller distances. 

Figure \ref{fig:dist_study}(c) shows the values for the proximity induced potential $\Delta$ and 
the Rashba spin-orbit parameter $\lambda_\text{R}$. The Rashba parameter is increasing steadily 
with decreasing distance. We also plot the derivative of the Rashba parameter with respect to the 
distance $-\partial \lambda_\text{R}/\partial d_z$. One sees that the Fermi level 
shift and the change in the Rashba parameter are correlated by comparing the derivative of the 
Rashba parameter to the Fermi level shift. Both curves change their trend at 2.5~\AA. We can see that the origin of the Rashba spin-orbit 
coupling is due to charge doping (determined by the Fermi energy shift $E_\textrm{D}$), leading to a built-in electric field, and
due to the positioning of the graphene sheet in the electrostatic potential of the Cu(111) surface. At the distance of 
2.5~\AA\, the charge doping stops, and therefore the Rashba spin-orbit coupling increases at a 
lower pace. It remains increasing though, as the graphene sheet resides in a potential which becomes 
steeper as it gets closer to the nuclei of copper.
It is surprising that the $\Delta$, which first increases from larger 
to smaller distances, decreases, becomes zero at 2.4~\AA\, and then inverts its sign. We will discuss this
in more detail later.
We estimate the pressure $p$ one would have to exert on graphene to reach this distance as
\begin{align*}
 p &= \frac{\Delta E}{\Delta d_z\cdot A} = \frac{200~\text{meV}}{(3.09-2.40)~\text{\AA}\cdot 
(2.46~\text{\AA})^2 \cdot \sin{60^\circ}} \\
   &= 8.8~\text{GPa},
\end{align*}
where $\Delta E$ is the energy difference between the lowest energetic state and the state where the 
transition happens, $\Delta d_z$ their distance difference, and $A$ is the area of the unit 
cell. The bulk modulus of copper for comparison is 184~GPa.\cite{Wyckoff1976}

The amplitudes of the intrinsic spin-orbit coupling parameters $\lambda_\text{I}^\text{A}$ and $\lambda_\text{I}^\text{B}$
strongly increase as graphene is pushed towards the Cu(111) surface, see 
Fig.~\ref{fig:dist_study}(d). For large distances both parameters tend to values comparable in size as in pure graphene.
For smaller distances the sublattice asymmetry transfers to the parameters and $\LIA$ is much
stronger affected due to the specific graphene sublattice positioning on Cu(111). $\LIA$ 
reaches values up to 7~meV, whereas $\lambda_\text{I}^\text{B}$ stays smaller than 1~meV for all tested
distances and tends to saturate at 1~meV when reaching a small distance of 1.8 \AA.
\begin{figure}
 \centering
 \includegraphics[width=0.3\textwidth]{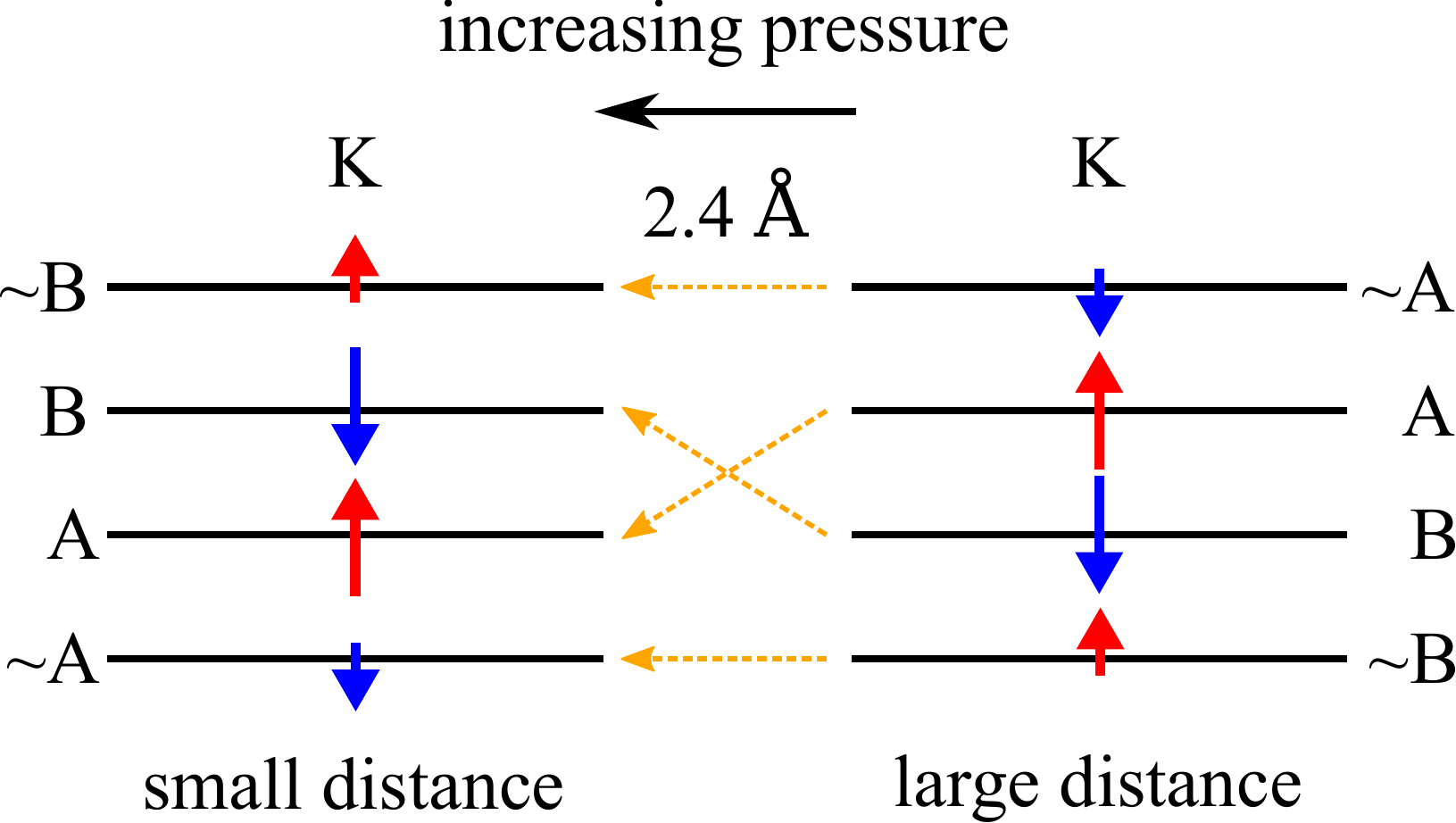}
 \caption{(color online) Scheme visualizing the transition of spin states at K with vertical pressure. 
Black solid lines indicate the energy levels, A and B stands for the sublattice. 
Arrows pointing upwards (downwards) represent spins pointing along $z$ ($-z$), 
shorter arrows indicate spin mixture and their projection to the $z$ direction.
}
 \label{fig:transition_scheme}
\end{figure}

The last two panels in Fig.~\ref{fig:dist_study}(e) and (f) show the spin-$z$ expectation 
values at K for the eigenvalues $\varepsilon_i$, where 1 labels the lowest energy and 4 the highest 
energy state. From Fig.~\ref{fig:dist_study}(e) we can see that the outermost expectation 
values represent spin states of mixed spin, as values of spin 1/2 are only reached, when graphene 
is well separated from copper. The spin expectation values of states 2 and 3 are pure states and are
always quantized in $z$ direction. When the hybridization gap closes, at 2.4~${\textrm\AA}$, 
we observe, that the signs of all spin expectation values change abruptly. This behavior is exemplified 
in Fig.~\ref{fig:transition_scheme}. When the distance of graphene to copper is decreased, 
the spin as well as pseudospin signs change.

\begin{figure}
 \centering
 \includegraphics[width=0.45\textwidth]{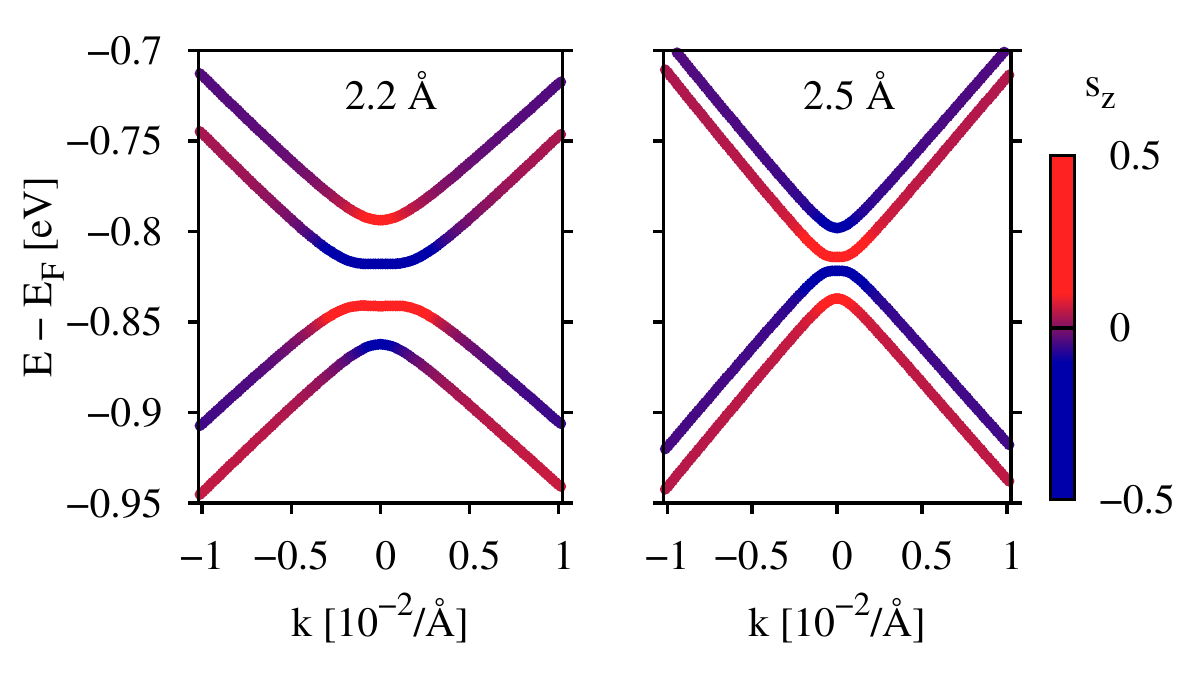}
 \caption{(color online) Band structure topologies of graphene on Cu(111) 
for 2.2~\AA\, and 2.5~\AA\, distances of graphene from Cu(111) surface.
The spin $s_z$ expectation values for the states are encoded by the color scale, where red (gray) color
denotes spin-$z$ expectation value of 1/2 and blue (black) color denotes a spin-$z$ expectation value of -1/2. The color scale is set such that $\pm 0.1$ of $s_z$ leads to a saturated color.
In this way, trends of how spin expectation values evolve in the bands are better visible.}
 \label{fig:band_structure_topology}
\end{figure}
In Fig.~\ref{fig:band_structure_topology} we show the topology of the bands obtained from DFT calculations around K for distances 
of 2.2 and 2.5~${\textrm \AA}$, with the corresponding spin-$z$ expectation values.
 The plot is consistent with 
Fig.~\ref{fig:dist_study}(e) and (f), for 2.5~\AA\, the band structure resembles the one in 
Fig.~\ref{fig:bands_fit} and has spin up-down-up-down sequence, where the inner eigenstates 
have pure $s_z= \pm 1/2$ components. The spin-$z$ character within the bands stays the same. 
The band structure topology for 2.2~\AA\, is different. 
At the K point the inner eigenstates again have pure $s_z=\mp1/2$ spin, but all signs are reversed.
Furthermore, the spin-$z$ character is not preserved within the bands. There is evidence for a band inversion for the inner bands
with a significant spin mixing to outermost bands. The spin reversal is accompanied by a change of the pseudospin character of the states.
The valence states become localized on the A sublattice and conduction bands on sublattice B. We note that similarly to the spin mixing for the outermost
bands, the states are also sublattice mixed, which is also depicted in Fig.~\ref{fig:transition_scheme}.
Our model is able to reproduce the spin-$z$ behavior of Fig. \ref{fig:band_structure_topology} (not shown here).


\section{\label{sec:concl} Summary}
We have shown that the electronic band structure measured by ARPES is reasonably described by DFT+$U$ calculations. We are able to correctly describe the Fermi level position
and copper $d$ band onset. Based on this good orbital description we predict the spin-orbit coupling effects by analyzing the low energy graphene states using a robust model Hamiltonian. We show, that our Hamiltonian is able to describe the spin-orbit induced
band splittings even away from the K point.
We extracted spin-orbit coupling parameters as well as spin expectation values dependent on
the graphene\textendash copper distance and found a strong distance-dependent behavior of spin-orbit coupling parameters and a reordering of 
the spin and pseudospin structure at the Dirac point at $d_z$=2.4~\AA. At low distances the Dirac band structure gets inverted due to the overlap of opposite spin valence and conduction bands. Our findings are experimentally verifiable with techniques such as ARPES, by increasing the resolution to resolve the meV and sub meV spectral ranges. 

\begin{acknowledgments}
This work was supported by the DFG SFB Grant No. 689 and GRK Grant No. 1570, and by the EU Seventh Framework Programme under Grant Agreement No. 604391 Graphene Flagship.
\end{acknowledgments}


\bibliography{paper}

\end{document}